\documentclass[preprint]{aastex}
\usepackage{epsfig}
\usepackage{times}

\shorttitle{A Search for Disks in TWA}
\shortauthors{Weinberger et al.}

\newcommand{\lir}{L$\rm _{IR} / L_\star$}
\newcommand{\lprime}{L$^\prime$}

\begin{document}

\title{A Search for Warm Circumstellar Disks in the TW Hydrae
Association}

\affil{\bf\small{(Accepted for Publication in Astronomical Journal, April 2004 issue)}}

\author{A. J. Weinberger}
\affil{Department of Terrestrial Magnetism, Carnegie Institution of
Washington,\\ 
and NASA Astrobiology Institute\\
5241 Broad Branch Road, NW, Washington, DC 20015}

\email{weinberger@dtm.ciw.edu}

\author{E. E. Becklin, B. Zuckerman, and I. Song.}
\affil{Division of Astronomy, University of California Los Angeles\\
and NASA Astrobiology Institute\\
Box 951562, Los Angeles, CA 90095-1562}

\email{becklin,ben,song@astro.ucla.edu}

\begin{abstract}

A search for previously undetected optically thin disks around stars in
the nearby, young, TW Hydrae Association was conducted around sixteen
stars with sensitive 12 and 18~\micron\ photometry.  The survey could
detect Zodiacal-like dust, with temperature 200--300 K, at levels of
\lir=7$\times 10^{-3}$. Possible mid-infrared excess emission from TWA
17 was detected at the 2$\sigma$ level, but none of the other stars
showed evidence for circumstellar dust.  The rapid disappearance of
large amounts of dust around the K and M-type stars in this sample may
mean that any planet formation in the terrestrial planet region was
completed very quickly. There appears to be a bi-modal distribution of
dust disks in TWA with stars having either copious or negligible warm
dust.

\end{abstract}


\section{Introduction}

The TW Hydrae Association (TWA) provides an excellent laboratory in
which to study the evolution of dusty disks.  This collection of 24 star
systems within $\sim$40~pc of each other
\citep{delareza89,Webb99,Sterzik99,Webb00,Zuckerman01,Song03} is one of
the the closest ($\sim$60 pc) young stellar groups to Earth.
Primary members range in spectral type from A0 to M3 and have common
ages of 5--10 Myr as determined from pre-main sequence tracks, Li
abundances and X-ray fluxes. Thus TWA stars are poised between T Tauri and
main sequence stages of stellar evolution.

Inner (radius $<$ 0.1 AU) disks around weak-lined T Tauri stars are
observed to dissipate very quickly, in $\lesssim$6 Myr, and
contemporaneously with the cessation of accretion \citep{Haisch01}. For
planet formation, it is of interest how quickly material dissipates
further out in the disk, so observations at longer wavelengths sensitive
to colder dust are necessary.  Results from ISO \citep{Spangler,Meyer00}
again suggest that disks at 0.3-20 AU disappear quickly.

Dusty disks around four TWA stars were inferred from excess infrared
emission discovered with the Infrared Astronomical Satellite (IRAS), and
they are strikingly diverse. The A0 star HR 4796A (TWA 11) has a narrow
ring of dust located at 70 AU
\citep{Jaya98,Koerner98,Schneider99,Telesco00}. TW Hya (TWA 1) has a
broad face-on disk that extends more than 135 AU from the star with a
dip in surface brightness at $\sim$110 AU
\citep{Krist00,Wilner00,Weinberger02}. TW Hya's disk still contains much
molecular gas \citep{Kastner97} while HR 4796A's has no detectable gas
\citep{Greaves00}.  IRAS also measured excess emission from Hen 3-600
and HD 98800, which are both multiple stars where the dust responsible
for the infrared excess encircles one member \citep{Jaya99a,Gehrz99}
while cooler dust detected at submillimeter wavelengths may be in
circumbinary orbits \citep{ZuckermanARAA}.  The amount of dust,
characterized by the fraction of the star's luminosity which is
re-radiated in the mid-to-far-infrared ($f$), is 5$\times10^{-3}$ for HR
4796A and much larger for the other three stars ($f$=0.1--0.3;
\citet{ZFK95}).

The IRAS Faint Source Catalog, complete (SNR$\approx$5) to a flux
density limit of $\sim$0.2 Jy at 12 and 25~\micron, had insufficient
sensitivity to probe fully for disks in TWA.  At the distance to and age
of TWA, 12~\micron\ stellar photospheric emission was detectable by IRAS
only for spectral types earlier than K0.  We have performed a ground based
search with better sensitivity to search for warm circumstellar disks
that might lurk around sixteen TWA stars.

\section{Observations, Data Analysis, and Photometry}

Observations were made at the W. M. Keck Observatory using the facility
instrument Long Wavelength Spectrometer (LWS) \citep{jones}. LWS uses a
$128\times128$\ pixel Boeing Si:As detector, and has a plate scale of
0.080 arcsec pixel$^{-1}$, resulting in a focal-plane field of view of
$10.2\arcsec$ square.  The characteristics of the filters used in this
project are given in Table \ref{tab_filters}.  \lprime\ band (3.8 $\mu$m)
measurements were obtained for two purposes: (1) quickly locating the star
on the array and (2) providing a long wavelength point for calculations of
stellar photospheric fluxes.  We used the most sensitive 12 and 18~\micron\
filters available during each run; the SiC and 17.65 filters were
added in 2001 and so were not available at the start of this project.
TWA 5--25 are all identified as members of TWA in \citet{Webb99},
\citet{Sterzik99}, \citet{Zuckerman01}, or \citet{Song03}.

A summary of the photometric properties of the six nights used for
observations is given in Table \ref{tab_nights}.  All six nights were
clear of clouds. During each night, infrared standard stars were
observed approximately every two hours and a photometric calibration for
each night of observations was obtained using all standards from that
night.  Sensitivities were characterized in terms of an instrumental
magnitude (mag $-2.5 \,\rm{log}(adu/s)$) and are given in Table
\ref{tab_nights}. These standards were usually at small airmass (1--1.6)
compared to the TWA targets, but on two nights -- 6 Feb 2001 and 29 May
2003, standards were observed at airmasses up to 3.3 and used to
determine airmass corrections in units of mag airmass$^{-1}$. Standard
deviations of measurements of multiple standards and linear fits to the
sensitivity as a function of airmass were used to determine the
calibration uncertainty.  This was taken to be the systematic error in
the photometry of the target stars.

We employed a standard chop-nod scheme for background removal, in which
the target star image falls on the detector in either two or all four
observed positions.  The telescope's secondary mirror was chopped
5--10\arcsec\ north-south at frequencies of $\sim$5 Hz, and images from
each chop position were summed.  After $\sim$30 s, the telescope was
nodded the same distance as the chop and images were again summed in
both chop positions. The double difference ($(nod_1 chop_1 - nod_1
chop_2) - (nod_2 chop_1 - nod_2 chop_2)$) was used to remove the thermal
background from the sky and telescope.  Total on-star integration times
are given in Tables \ref{tab_photoL}--\ref{tab_photoQ}. 

Different LWS filters place images at different locations on the
detector, so standard stars were used to determine the shifts between
the \lprime, 12, and 18~\micron\ filters.  For filters in which the
targets were not visible at high SNR, the predetermined shifts were
used to locate the target.  The uncertainty in the prediction is about
one pixel, probably due to offset guiding errors.

The seeing at airmass 1.5 was 0.8\arcsec\ at \lprime-band and 0.5\arcsec
at 12 and 18~\micron\ for 2000 and 2001 runs and 0.5\arcsec at
\lprime-band and 0.4\arcsec\ at 12 and 18~\micron\ during the 2002 run.
The image size at higher airmass was measured on 6 Feb 2001 and 29 May
2002 and was consistent with seeing proportional to $({\rm cos}
z)^{0.6}$.

The flux density of each star was measured with synthetic aperture
photometry on the double differenced images in a beam of radius 14
pixels ($\sim$1$''$).  These measurements were corrected to the total
flux by measuring an aperture correction from the standard
stars. Imperfect background subtraction was accounted for by subtracting
a ``sky'' value measured in an annulus around the star.  Although, in
principle, higher SNR photometry is obtained with an aperture radius
approximately equal to the FWHM of the stars, we used a larger beam so
we would be insensitive to the uncertainty in the location of the object
in the filters where it was not seen at high SNR.  The photometric
uncertainties reported are a combination of the sky noise, the Poisson
noise from the stellar measurement, and the systematic calibration
uncertainty described above.  We estimate at most an additional 2\%
uncertainty from pixel-pixel sensitivity variations across the array;
the images were not flat-fielded.

\section{Results}

Photometric measurements of the sixteen target stars are given in
Tables \ref{tab_photoL}-\ref{tab_photoQ}.  All but three were detected
at 12~\micron\ with SNR$>$3. None of the stars was detected with
SNR$>$3 at 18~\micron.

\subsection{Assessing the Presence of Infrared Excess\label{nextgen}}

We fit stellar photosphere models \citep{NextGen} to J, H, and Ks-band
data from the 2MASS catalog and our \lprime-band data where available.  We
treated the stellar effective temperature and the model normalization as
unknowns and performed a chi-squared minimization to determine these
parameters.  The models have temperature bins of 100~K in the region
1700--4000 K and 200~K in the region 4000--10000~K.  The best fit models
were used to predict the photospheric flux densities at 12 and 18 $\mu$m
that are reported in Tables \ref{tab_photoN} and \ref{tab_photoQ}.

It is not obvious how to assess the uncertainties in the photospheric
model predictions other than by comparison with measurements of stars
without excess. As a test, instead of using the best fit temperatures,
we used the \citet{Hartigan94} and \citet{Luhman98} temperature scales
for dwarfs to convert published spectral types to effective temperatures
and compared the resulting predictions.  On average, the predictions
were the same, i.e. there was no bias. The scatter in the difference of
the predictions was 15\%.  We take this to be an upper limit on the
uncertainty in our knowledge of the photospheres because it could be
affected by incorrect spectral typing.  

\begin{figure}[htb]
\epsfig{file=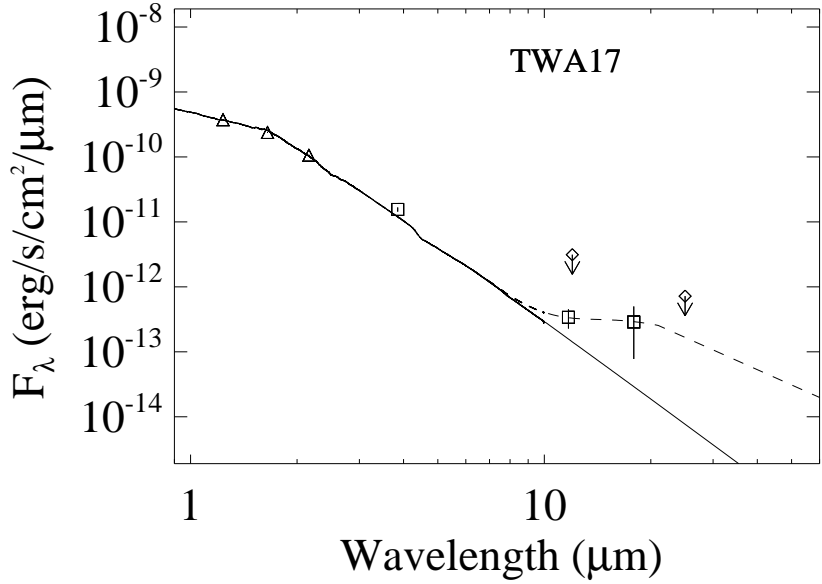,width=5in,clip=}
\figcaption[Weinberger.f1.ps]{Spectral energy distribution of TWA
17. Triangles are J, H, and Ks data from 2MASS, squares are the LWS data
reported in this paper, diamonds are the 150 mJy completeness level of
IRAS at 12 $\mu$m and extrapolated to 18$\mu$m.  The dotted line shows a
170 K blackbody, the best fit (i.e. color temperature) of the possible
excess we measure, corresponding to \lir = 0.005.\label{fig_twa17sed}}
\end{figure}

At 12 and 18 $\mu$m, all of our measurements agree to within 3 $\sigma$
with the model predictions.  In one object, TWA 17, both the 12 and 18
$\mu$m flux densities exceed the photospheric predictions, as seen in
Figure \ref{fig_twa17sed}. Using the total uncertainties on both points,
its excess is significant only at 2.2$\sigma$. The color temperature of
this excess is 170~K and corresponds to 0.005 L$_*$.  Three of the stars
in our sample were previously observed by \citet{Jaya99b} -- TWA 5A, TWA
6, and TWA 7.  The results for TWA 6 and 7 agree within
1$\sigma$. However, the measurement reported here for TWA 5A at 12
$\mu$m, 63.4 $\pm$ 7.4 mJy differs by 4$\sigma$ from their N-band (8--13
$\mu$m) measurement of 96 $\pm$ 9 mJy. Therefore, we do not confirm the
excess they suggested around TWA 5A.

\begin{figure}[htb]
\epsfig{file=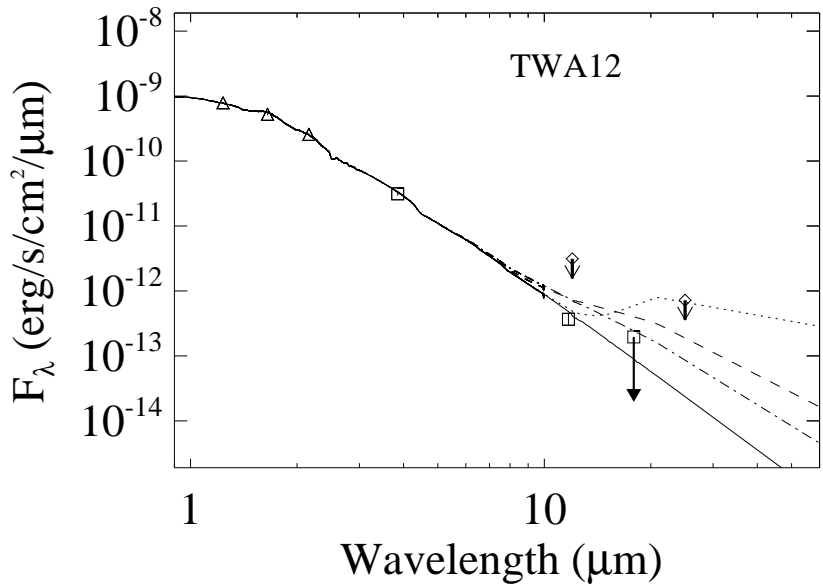,width=5in,clip=}
\figcaption[Weinberger.f2.ps]{Spectral energy distribution of TWA 12
presented as an example of our excess limits.  Triangles are J, H, and
Ks data from 2MASS, squares are the LWS data reported in this paper,
diamonds are the 150 mJy completeness level of IRAS at 12 $\mu$m and
extrapolated to 18$\mu$m.  The dotted, dashed, and dotted-dashed lines
are the maximum 100 K, 200 K, and 300 K blackbodies of infrared excess
that can be accommodated by our and the IRAS measurements at 3
$\sigma$.\label{fig_twa12sed}}
\end{figure}

To quantify the limits on the amount of excess emission that could be
present around each star, we found the maximum luminosity blackbodies
consistent with the 3 $\sigma$ upper limits of the measured 12 and 18
$\mu$m flux densities. Real disks are likely to have dust at a range of
temperatures, and we do not know the disk energy distributions a
priori. So we take the simplest case of blackbodies at three
temperatures -- 100, 200, and 300K.  We characterize the upper limit on
the excess at each of these temperatures by L$_{\rm BB}$/L$_\star$ and
report these values in Table \ref{tab_taus}.  A sample set of fits is
shown in Figure \ref{fig_twa12sed} for TWA 12. Given our sensitivities
at 12 and 18 $\mu$m, we are most sensitive to dust at 200-300K. For
comparison, HR 4796 has \lir=5$\times$10$^{-3}$ at T=110~K
\citep{Jura93}, although for this A-type star its dust is at Kuiper-belt
like distances.  For 300~K dust, in the terrestrial planet region around
the sample stars, we could detect \lir=5$\times$10$^{-3}$ for 9 of the
16 TWA stars.

\section{Discussion}

Most previous surveys of post T-Tauri star dust have concentrated on
$\sim$1000 K dust, observable at L-band.  This dust resides at
tens of stellar radii only. One set of surveys show that these inner
edges disappear at 3--6 Myr, shortly after accretion ends
\citep{Haisch01}.  A survey of the mostly very low mass members of the
$\eta$ Cha cluster, however, finds a high hot disk fraction even at an
age of 5--8 Myr \citep{Lyo03}.  Whether these differences stem from sample
selection effects, birth conditions or some other variables remains to
be determined.

Our measurements constrain the amount of dust at temperatures of
$\sim$200 K. For the TWA stars, this corresponds to distances of 1--5
AU, depending on the luminosity of the star and the emissivity of the
grains.  The median spectral type of our sample is M0 with a luminosity
of $\sim$0.25 L$_\odot$. Grains that absorb and emit like blackbodies
reach a temperature of 200 K at 1 AU for such a star while interstellar
silicates at 200 K are at 4 AU.

Converting the infrared excess limits to dust masses is hampered by our
lack of knowledge of the dust grain size distribution.  For example, if
all the grains in a disk had radius 20 $\mu$m, approximately the size
our observations are most sensitive to, had a semi-major axis of 1 AU,
and total \lir = 0.01, their mass would be $\sim$10$^{-5}$
M$_{Earth}$. However, if the number of grains of a given radius followed
a power-law distribution with slope -3.5 and a maximum size of 1000 km,
the dust mass would be 5 orders of magnitude larger.

Dust around young stars could come in two forms: primordial material,
largely unprocessed from the interstellar medium, or debris generated in
collisions of planetesimals.  The material around at least one member of
TWA, HR 4796A, is likely debris \citep{Jura95,Schneider99} based on its
inferred size, color, and albedo.  So, planetesimals evidently have had
time to form around this A0 star.  The dust around TW Hya itself, a K7
star, is likely a mixture of primordial and debris based on the grain
size distribution which is larger than interstellar
\citep{Weinberger02}.

Primordial, i.e. interstellar, dust distributions, are dominated by
small ($\lesssim$1 $\mu$m) grains. Our observations would be very
sensitive to primordial dust, and therefore it is likely that the stars
for which we obtain upper limits on excesses have already lost
such material.

Models of planetesimal formation produce copious quantities of dust
during an epoch where large bodies perturb each other and generate many
collisions \citep[e.g.][]{Kenyon02}.  In our own Solar System, collisions
must have occurred in the terrestrial planet region, at least
at a low level, for 30 Myr, the likely time of the Moon forming impact
\citep{Yin}.  

Given the low dust limits set by our observations of TWA members, the
era of frequent collisions must have stopped at 1--4 AU long enough ago
for little dust to remain at temperatures of $\sim$200 K.  In models of
terrestrial accumulation, dust formation declines exponentially over
1--2 Myr after peaking during the formation of 1000-2000 km bodies
\citep{Kenyon04}. By 3 Myr after this period, they predict an \lir\ of
$<$1e-3. Roughly then, the epoch of large planetesimal growth must have
ended about 1 Myr ago for the dust to have disappeared.  Thus, the
absence of material in the terrestrial planet region implied by the
non-detections of our survey may mean that planet formation has already
taken place.  Some evidence for rapid planet formation may in fact exist
in the structure of the debris around HR 4796A which is confined to a
very narrow ring with brightness asymmetry \citep{Schneider03}.

It appears from radial velocity surveys that giant planets commonly
orbit main sequence Solar type stars, but less frequently orbit the
M-type stars that comprise most of our sample \citep{Gould03}. Only one
star in the TW Hydrae association, TW Hya itself, is a single star with
large amounts of dust and gas remaining.  Given the current statistics,
it is possible that planets do form only in the long lived disks like
that surrounding TW Hya.  However, the number of known planets is likely
to grow as lower masses and longer orbits are probed by experiments, and
this study shows that there are no other long lived disks in the TW Hya
association that can still form planets.

\section{Conclusions}

This survey could detect small amounts of dust, similar to that found
around the well studied TW Hya Association A-type star HR 4796A, around
the late-type members of the association.  No new stars with
infrared excess at 12 or 18 $\mu$m were positively discovered in this
most sensitive survey to date.  It is remarkable that of the 24 star
systems now known in the TW Hydrae association, all of those with dust
were discovered by IRAS despite the ten times better sensitivity of this
survey.  Dust content in TWA is bimodal, with a few stars possessing
much dust and most largely devoid of material. In regions analogous to
the terrestrial planet region, even at the young age of 5--10 Myr,
little material remains or is being generated in collisions around these
late-type stars.

\acknowledgements

This paper is based on observations at the W. M. Keck Observatory, which
is operated as a scientific partnership between the California Institute
of Technology, the University of California, and NASA, and was made
possible by the generous financial support of the W. M. Keck
Foundation. We recognize the importance of Mauna Kea to Hawaiians. We
received valuable assistance with observations from Randy Campbell, Meg
Whittle, Joel Aycock, and Ron Quick.  We acknowledge support from NASA
Origins of Solar Systems grants to UCLA and CIW. This publication makes
use of data products from the Two Micron All Sky Survey funded by NASA
and NSF.  We thank Anna Haugsjaa, an NSF Research Experience for
Undergraduates summer intern at CIW for her work on this project.

\begin{deluxetable}{lccc}
\tablecaption{Filter parameters\label{tab_filters}}
\tablenum{1}
\tablewidth{0pt}
\tablehead{
\colhead{Name}
        &\colhead{Central}
                  &\colhead{Bandpass}
                             &\colhead{Zero mag}\\
\colhead{}
        &\colhead{Wavelength}
                  &\colhead{}
                             &\colhead{Flux Density}\\
\colhead{} &\colhead{($\mu$m)} 
	          &\colhead{($\mu$m)}
                             &\colhead{(Jy)}}
\startdata
\lprime &3.85     &3.5--4.2      &251.0 \\
11.7    &11.67    &11.2--12.2    &29.9  \\
SiC     &11.77    &10.5--12.9    &29.6  \\
17.65   &17.74    &17.3--18.2    &12.9  \\
17.9    &17.9     &16.9--18.9    &12.7  \\
\enddata
\end{deluxetable}

\begin{deluxetable}{lccccccl}
\tablenum{2}
\tabletypesize{\scriptsize}
\tablecaption{Nightly Photometric Sensitivities\label{tab_nights}}
\tablewidth{0pt}
\tablehead{
\colhead{Date of}
            &\colhead{\lprime}
                        &\colhead{\lprime\ air mass}
                                &\colhead{N-band}
                                                  &\colhead{N air mass}
					                &\colhead{Q-band}
						                          &\colhead{Q air mass}
                                                                                  &\colhead{}\\
\colhead{Observation}
            &\colhead{Zero point}
                        &\colhead{correction}
                                &\colhead{Zero point}
                                                  &\colhead{correction}
					                &\colhead{Zero point}
						                          &\colhead{correction}
                                                                                  &\colhead{Stars observed}
}
\startdata
2000 Dec 11 &\nodata &\nodata  &13.56 $\pm$ 0.07 &0.27 &11.33 $\pm$ 0.04    &0.31 &TWA 6\\
2001 Feb 05 &\nodata &\nodata  &13.46 $\pm$ 0.09 &\nodata &11.03 $\pm$ 0.11 &0.29 &TWA 7, 16\\
2001 Feb 06 &15.34   &0.03     &13.49 $\pm$ 0.09 &0.09 &11.31 $\pm$ 0.09    &0.39 &TWA 5, 12, 13, 15, 17\\
2002 May 28 &\nodata &\nodata  &14.70            &\nodata &11.38            &\nodata &TWA 14, 18, 25\\
2002 May 29 &15.79   &\nodata  &14.71 $\pm$ 0.04 &0.18 &11.25 $\pm$ 0.18    &0.54 &TWA 19\\
2002 Jun 01 &15.74   &\nodata  &14.69            &\nodata &11.29            &\nodata &TWA 23\\
\enddata

\tablecomments{Zero points are given as instrumental magnitudes and
airmass corrections are given as magnitudes per unit airmass. The N and
Q-band values listed correspond to the SiC and 17.65 filters on 2002 May
28 and Jun 01 and for the 11.7 and 17.9 $\mu$m filters otherwise.}

\end{deluxetable}

\begin{deluxetable}{lllllll}
\tablenum{3}
\tabletypesize{\footnotesize}
\tablecaption{Target Star Properties and \lprime-band Results\label{tab_photoL}}
\tablewidth{0pt}
\tablehead{
\colhead{Star}
          &\colhead{Sp. Type}
               &\colhead{Airmass}
                     &\colhead{time}
                             &\colhead{Flux Density}
                                     &\colhead{Statistical Unc.}
                                             &\colhead{Total Unc.}\\
\colhead{} &\colhead{}&\colhead{(s)} &\colhead{(mJy)}
			&\colhead{(mJy)}&\colhead{(mJy)}
}                        
\startdata
TWA 5       &M1.5   &1.75      &48     &555.7  &4.4  &28.4  \\
TWA 6       &K7     &1.73      &30     &217.2  &5.8  &12.4 \\
TWA 12      &M2     &2.30      &192    &154.5  &2.1  &57  \\
TWA 13N     &M2     &1.72      &96     &262.3  &3.0  &9.3  \\
TWA 13S     &M1     &1.72      &96     &270.2  &3.4  &14.0  \\
TWA 14      &M0     &2.46      &45     &118.9  &4.1  &8.4  \\
TWA 15A     &M1.5   &2.68      &96     & 35.4  &3.1  &3.9  \\
TWA 15B     &M2     &2.68      &96     & 38.0  &3.1  &4.0  \\
TWA 17      &K5     &2.45      &96     & 76.8  &3.2  &5.7  \\
TWA 18      &M0.5   &2.29      &30     & 80.9  &4.7  &6.6  \\
TWA 19A     &G5     &2.86      &30     &230.1  &4.9  &16.6  \\
TWA 19B     &K7     &2.89      &30     &124.0  &4.9  &9.9  \\
TWA 23      &M1     &1.66      &30     &238.9  &5.2  &12.9  \\
TWA 25      &M0     &2.00      &60     &327.7  &4.8  &18.2  \\
\enddata
\end{deluxetable}

\begin{deluxetable}{lccccccc}
\tablenum{4}
\tabletypesize{\footnotesize}
\tablecaption{12 $\mu$m Results\label{tab_photoN}}
\tablewidth{0pt}
\tablehead{
\colhead{Star}
          &\colhead{Airmass}
	          &\colhead{filter}
                          &\colhead{time}
                             &\colhead{Flux Density}
                                     &\colhead{Statistical Unc.}
                                             &\colhead{Total Unc.}
							 &\colhead{Prediction}\\
\colhead{} &\colhead{} &\colhead{}&\colhead{(s)} 
                 &\colhead{(mJy)}&\colhead{(mJy)}&\colhead{(mJy)} &\colhead{(mJy)}
}             
\startdata
TWA 5      &1.86  &11.7    &192    &63.4   &4.5 &7.4  &63.1\\
TWA 6      &1.64  & 11.7   &120    &20.3   &5.5 &5.5  &29.5\\
TWA 7      &1.91  & 11.7   &120    &70.4   &5.6 &8.6  &76.4\\
TWA 12     &2.56  & 11.7   &384    &16.7   &3.0 &3.5  &17.5\\
TWA 13N    &1.72  &11.7    &192    &36.5   &4.7 &5.7  &29.8\\
TWA 13S    &1.72  &11.7    &192    &36.0   &4.3 &5.4  &30.7\\
TWA 14     &2.59  &SiC     &216    &12.4   &3.8 &3.8  &13.5\\
TWA 15A    &2.67  &11.7    &240    & 4.2   &5.1 &5.1  & 4.0\\
TWA 15B    &2.67  &11.7    &240    &-1.7   &5.2 &5.2  & 4.3\\
TWA 16     &2.40  &11.7    &192    &22.1   &4.6 &5.2  &27.7\\
TWA 17     &2.45  &11.7    &192    &15.5   &4.8 &5.1  & 8.7\\
TWA 18     &2.29  &SiC     &216    &10.8   &3.5 &3.5  & 9.2\\
TWA 19     &2.87  &11.7    &120    &18.6   &3.9 &3.9  &26.1\\
TWA 19B    &2.90  &11.7    &216    &11.1   &2.7 &2.7  &14.1\\
TWA 23     &1.67  &SiC     &216    &27.4   &2.9 &3.1  &27.1\\
TWA 25     &2.01  &SiC     &216    &41.9   &4.2 &4.5  &37.2\\
\enddata
\end{deluxetable}

\begin{deluxetable}{lccccccc}
\tablenum{5}
\tablecaption{18 $\mu$m Results\label{tab_photoQ}}
\tabletypesize{\footnotesize}
\tablewidth{0pt}
\tablehead{
\colhead{Star}
          &\colhead{Airmass}
	          &\colhead{filter}
                          &\colhead{time}
                             &\colhead{Flux Density}
                                     &\colhead{Statistical Unc.}
                                             &\colhead{Total Unc.}
							 &\colhead{Prediction}\\
\colhead{} &\colhead{} &\colhead{}&\colhead{(s)} &\colhead{(mJy)}&\colhead{(mJy)}&\colhead{(mJy)}&\colhead{(mJy)}
}
\startdata
TWA 5      &1.83  & 17.9   &618    & 27.4  &16.4   &16.6  &16.6\\
TWA 6      &1.71  & 17.9   &983    &  7.1  &12.6   &12.6  &12.6\\
TWA 7      &1.95  & 17.9   &319    &-10.9  &31.5   &31.6  &31.6\\
TWA 12     &2.24  & 17.9   &618    & 17.2  &20.8   &20.9  &20.9\\
TWA 13N    &1.73  &17.9    &618    & -0.3  &17.1   &17.1  &17.1\\
TWA 13S    &1.73  &17.9    &618    & 11.8  &18.5   &18.5  &18.5\\
TWA 14     &2.48  &17.65   &408    & 29.7  &38.6   &39.1  &39.1\\
TWA 15A    &2.67  &17.9    &618    &-16.3  &25.4   &25.5  &25.5\\
TWA 15B    &2.67  &17.9    &618    & 21.6  &24.7   &24.8  &24.8\\
TWA 16     &2.42  &17.9    &319    & 41.7  &40.3   &41.1  &41.1\\
TWA 17     &2.44  &17.9    &618    & 30.8  &22.3   &22.5  &22.5\\
TWA 18     &2.31  &17.65   &408    & 46.8  &32.2   &33.5  &33.5\\
TWA 23     &1.69  &17.65   &408    & 22.3  &25.6   &25.9  &25.9\\
TWA 25     &2.03  &17.65   &408    &-19.3  &29.9   &30.1  &30.1\\
\enddata
\end{deluxetable}

\begin{deluxetable}{lccc}
\tablenum{6}
\tablecaption{3$\sigma$ \lir\  ($\times 10^{-3}$) limits for disk detection at
three dust temperatures \label{tab_taus}}
\tablewidth{0pt}
\tablehead{
\colhead{Star} &\colhead{100~K}  &\colhead{200~K}  &\colhead{300~K}
}
\startdata
TWA 5           & 4.0          & 1.1            & 1.0 \\
TWA 6           & 8.3          & 3.0            & 3.1 \\
TWA 7           & 6.4          & 1.8            & 1.7 \\
TWA 12          &15.8          & 3.6            & 3.0 \\
TWA 13N         & 6.8          & 2.5            & 2.7 \\
TWA 13S         & 7.4          & 2.5            & 2.6 \\
TWA 14          &50.9          &10.9            & 8.6 \\
TWA 15A         &98.7          &24.0            &20.5 \\
TWA 15B         &85.8          &21.1            &18.1 \\
TWA 16          &44.0          & 9.8            & 7.8 \\
TWA 17          &50.1          &14.2            &13.3 \\
TWA 18          &88.6          &19.1            &15.1 \\
TWA 19A         &15.8          & 2.4            & 1.4 \\
TWA 19B         &41.4          & 8.5            & 6.3 \\
TWA 23          &14.4          & 3.3            & 2.8 \\
TWA 25          &10.6          & 2.3            & 1.9 \\
\enddata
\end{deluxetable}

\begin{thebibliography}{7}

\bibitem[de la Reza et al.(1989)]{delareza89} de la Reza, R., Torres,
C.A.O., Quast, G., Castilho, B.V., \& Vieira, G.L. 1989, \apjl, 343, 61

\bibitem[Gehrz et al.(1999)]{Gehrz99} Gehrz, R.~D., Smith, N., 
Low, F.~J., Krautter, J., Nollenberg, J.~G., \& Jones, T.~J.\ 1999, \apjl, 
512, L55 

\bibitem[Gould, Ford, \& Fischer(2003)]{Gould03} Gould, A., 
Ford, E.~B., \& Fischer, D.~A.\ 2003, \apjl, 591, L155 

\bibitem[Greaves, Mannings, \& Holland(2000)]{Greaves00} Greaves, 
J.~S., Mannings, V., \& Holland, W.~S.\ 2000, Icarus, 143, 155 

\bibitem[Haisch, Lada, \& Lada(2001)]{Haisch01} Haisch, K.~E., 
Lada, E.~A., \& Lada, C.~J.\ 2001, \apjl, 553, L153 


\bibitem[Hartigan, Strom, \& Strom(1994)]{Hartigan94} Hartigan, 
P., Strom, K.~M., \& Strom, S.~E.\ 1994, \apj, 427, 961 

\bibitem[Hauschildt et al.(1999)]{NextGen} Hauschildt, P. H., Allard, F.,
Ferguson, J., Baron, E., \& Alexander, D. R. 1999, \apj, 525, 871

\bibitem[Jayawardhana et al.(1998)]{Jaya98}Jayawardhana, R., Fisher, S.,
Hartmann, L., Telesco, C., Pina, R.\ \& Fazio, G.\ 1998, \apjl, 503,
L79

\bibitem[Jayawardhana et al.(1999a)]{Jaya99a} Jayawardhana, R., 
Hartmann, L., Fazio, G., Fisher, R.~S., Telesco, C.~M., \& Pi{\~ n}a, 
R.~K.\ 1999a, \apjl, 520, L41 

\bibitem[Jayawardhana et al.(1999b)]{Jaya99b} Jayawardhana, R., 
Hartmann, L., Fazio, G., Fisher, R.~S., Telesco, C.~M., \& Pi{\~ n}a, 
R.~K.\ 1999b, \apjl, 521, L129 

\bibitem[Jones \& Puetter(1993)]{jones}Jones, B., \& Puetter, R. 1993,
in Proc. SPIE vol. 1946, 610

\bibitem[Jura et al.(1993)]{Jura93}Jura, M., Zuckerman, B., Becklin,
E.E. \& Smith, R.C. 1993, \apjl, 418, L37

\bibitem[Jura et al.(1995)]{Jura95} Jura, M., Ghez, A.~M., 
White, R.~J., McCarthy, D.~W., Smith, R.~C., \& Martin, P.~G.\ 1995, \apj, 
445, 451 


\bibitem[Kastner et al.(1997)]{Kastner97} Kastner, J. H., Zuckerman, B.,
Weintraub, D. A., \& Forveille, T. 1997, Science, 277, 67

\bibitem[Kenyon \& Bromley(2002)]{Kenyon02} Kenyon, S.~J.~\& 
Bromley, B.~C.\ 2002, \apjl, 577, L35 

\bibitem[Kenyon \& Bromley(2004)]{Kenyon04}  Kenyon, S.~J.~\& 
Bromley, B.~C.\ 2004, \apjl, submitted

\bibitem[Koerner et al.(1998)]{Koerner98}Koerner, D.\ W., Ressler, M.\
E., Werner, M.\ W.\ \& Backman, D.\ E.\ 1998, \apjl, 503, L83

\bibitem[Krist et al.\ (2000)]{Krist00} Krist, J.\ E., Stapelfeldt, K.\
R., M{\'e}nard, F.\, Padgett, D.\ L.\ \& Burrows, C.\ J.\ 2000, \apj,
538, 793

\bibitem[Low, Hines, \& Schneider(1999)]{Low99} Low, F.~J., 
Hines, D.~C., \& Schneider, G.\ 1999, \apjl, 520, L45 

\bibitem[Lowrance et al.(1999)]{Lowrance99} Lowrance, P.~J.~et 
al.\ 1999, \apjl, 512, L69 

\bibitem[Luhman \& Rieke(1998)]{Luhman98} Luhman, K.~L.~\& 
Rieke, G.~H.\ 1998, \apj, 497, 354 

\bibitem[Lyo et al.(2003)]{Lyo03} Lyo, A.-R., Lawson, W.~A., 
Mamajek, E.~E., Feigelson, E.~D., Sung, E., \& Crause, L.~A.\ 2003, \mnras, 
338, 616 

\bibitem[Meyer \& Beckwith(2000)]{Meyer00} Meyer, M.~R.~\& 
Beckwith, S.~V.~W.\ 2000, ISO Survey of a Dusty Universe, Proceedings of a 
Ringberg Workshop Held at Ringberg Castle, Tegernsee, Germany, 8-12 
November 1999, Edited by D.~Lemke, M.~Stickel, and K.~Wilke, Lecture Notes 
in Physics, vol.~548, p.341, 341 


\bibitem[Schneider et al.(1999)]{Schneider99}Schneider, G.\ et al.\
1999, \apjl, 513, L127 

\bibitem[Schneider et al.(2003)]{Schneider03}Schneider, G., Weinberger,
A.J., Smith, B.A., Becklin, E.E., Silverstone, M.J., Hines, D.C. \&
Lowrance, P.J. 2003, in preparation for AJ

\bibitem[Song, Zuckerman, \& Bessell(2003)]{Song03} Song, I., Zuckerman,
B., \& Bessell, M. S. 2003, ApJ, 599, in press (Dec 10 issue)

\bibitem[Spangler et al.(2001)]{Spangler} Spangler, C., Sargent, 
A.~I., Silverstone, M.~D., Becklin, E.~E., \& Zuckerman, B.\ 2001, \apj, 
555, 932 


\bibitem[Sterzik et al.(1999)]{Sterzik99} Sterzik, M. F., Alcala, J. M.,
Covino, E., \& Petr, M. G. 1999, \aap, 346, L41

\bibitem[Telesco et al.(2000)]{Telesco00}Telesco, C. M. et al.\ 2000,
ApJ, 530, 329

\bibitem[Webb et al.(1999)]{Webb99} Webb, R.~A., Zuckerman, 
B., Platais, I., Patience, J., White, R.~J., Schwartz, M.~J., \& McCarthy, 
C.\ 1999, \apjl, 512, L63 

\bibitem[Webb(2000)]{Webb00}Webb, R. A. 2000, PhD Thesis (UCLA)

\bibitem[Weinberger et al.(2002)]{Weinberger02} Weinberger, A. J., Becklin,
E. E., Schneider, G., Chiang, E. I., Lowrance, P. J., Silverstone, M.,
Zuckerman, B., Hines, D. C., \& Smith, B. A. 2002, \apj, 566, 409


\bibitem[Wilner et al.(2000)]{Wilner00}Wilner, D.\ J., Ho, P.\ T.\ P.,
Kastner, J.\ H.\ \& Rodr{\'i}guez, L .\ F.\ 2000, \apjl, 534, L101

\bibitem[Yin et al.(2002)]{Yin} Yin, Q., Jacobsen, S. B., Yamashita, K.,
Blichert-Toft, J., Telouk, P., \& Albarede, F. 2002, \nat, 418, 949

\bibitem[Zuckerman, Forveille, \& Kastner(1995)]{ZFK95} Zuckerman,
B. Forveille, T., \& Kastner, J.~H. 1995, \nat, 373, 494

\bibitem[Zuckerman(2001)]{ZuckermanARAA} Zuckerman, B. 2001, \araa, 39, 549

\bibitem[Zuckerman et al.(2001)]{Zuckerman01}
Zuckerman, B., Webb, R.~A., Schwartz, M., \& Becklin, E.~E.\ 2001, \apjl, 
549, L233 

\end{thebibliography}
\end{document}